\documentclass[twocolumn,showpacs,preprintnumbers,amsmath,amssymb,floatfix]{revtex4}

\usepackage{graphicx}
\usepackage{dcolumn}
\usepackage{bm}

\begin{document}


\title{
Electron coherent and incoherent
pairing
instabilities in inhomogeneous \\bipartite and nonbipartite
nanoclusters}

\author{A.~N.~Kocharian} 
\address{%
Department of Physics, California State University, Los Angeles,
CA 90032, USA; Physical Sciences, Santa Monica College, Santa
Monica, CA 90405, USA
}%
\author{G. W. Fernando and K.~Palandage}
\address{Department of Physics, University of Connecticut,
 Storrs, CT 06269, USA}%

\author{J.~W.~Davenport}
\address{Computational Science Center, Brookhaven National
Laboratory, Upton, NY 11973, USA}

\begin{abstract}
Exact calculations of collective excitations and charge/spin
(pseudo)gaps in an ensemble of bipartite and nonbipartite clusters
yield level crossing degeneracies, spin-charge separation,
condensation and recombination of electron charge and spin, driven
by interaction strength, inter-site couplings and temperature.
Near crossing degeneracies, the electron configurations of the
lowest energies control the physics of electronic pairing, phase
separation and magnetic transitions. Rigorous conditions are found
for the smooth and dramatic phase transitions with competing
stable and unstable inhomogeneities. Condensation of electron
charge and spin degrees at various temperatures offers a new
mechanism of pairing and a possible route to superconductivity in
inhomogeneous systems, different from the BCS scenario. Small
bipartite and frustrated clusters exhibit charge and spin
inhomogeneities in many respects typical for nano and
heterostructured materials. The calculated phase diagrams in
various geometries may be linked to atomic scale experiments in
high T$_c$ cuprates, manganites and other concentrated transition
metal oxides.

\end{abstract}
\maketitle
\section{Introduction}

Strongly correlated electrons in cuprates, manganites and other
transition metal oxides exhibit high T$_c$ superconductivity,
magnetism and ferroelectricity accompanied by spatial
inhomogeneities at the nanoscale level
~\cite{Davis1,Valla,Bodi,hashini,Boyer,Yazdani,Yazdani1,Cohen}.
Over the past few years there is an increase in interest to
electron instabilities in nanoclusters, assembled clusters of
correlated materials in various topologies for synthesizing new
nanomaterials with unique electronic and magnetic
properties~\cite{deHeer0,TM}. Obviously, there is a clear need for
an accurate analysis
 of electron
correlations, fluctuations and instabilities in nanoclusters and
large complex systems with competing phases. The closed form
solution, existing in the Bethe ansatz ground state~\cite{Yang},
is difficult to analyze at finite temperatures $T>0$ without
having to resort to various approximations. Perturbation theory is
usually inadequate while numerical methods have serious
limitations, such as in the Quantum Monte Carlo method with its
notorious sign problem where the resulting approximations often
lead to some controversy. On the contrary, exact calculations in
small clusters~\cite{Shiba,Falicov,Calaway,Schumann1,Hirsch} give
an appealing alternative for the detection of possible phase
separations and spatial inhomogeneities especially at finite
temperatures. As far as the authors are aware, an exact analysis
of level crossing instabilities (degeneracies) in canonical ground
state eigenvalues and corresponding competing average energies at
finite temperature for a general on-site interaction $U$ and
electron concentrations have not been attempted in small or
moderate size clusters~\cite{pairing}. Exact computations of
electron instabilities in various cluster geometries at the
nanoscale level can be vital to the understanding of the role of
thermal and quantum fluctuations for large pairing gaps and a
transition temperature T$_c$ in the correlated nanoclusters,
nanomaterials and corresponding ``large" inhomogeneous
systems~\cite{arXiv,JMMM,PRB,PLA,PRL,computer,cond-mat,to be
published}.

Although our approach for ``large" systems is only approximate,
this class of clusters in an ensemble displays a common behavior
which we believe is generic for large thermodynamic systems. Our
results for typical bipartite and frustrated (nonbipartite)
cluster geometries have successfully mapped out scenarios where
many body local effects are sufficient to describe spin-charge
separation and pairing pseudogaps at the nanoscale level. Spatial
microscopic inhomogeneities have been observed in a number of
scanning tunelling microscopy (STM) probes in doped high-Tc
superconductors (HTSCs). There is growing evidence suggesting that
"inhomogeneities" at the nanoscale level, in the so-called stripes
surrounded by essentially neutral correlated MH-like
antiferromagnetic insulators~\cite{Tranquada,Arigoni}, play a
defining role for the electron pairing and the origin of
superconductivity at the atomic scale in
HTSCs~\cite{Hudson,Balatsky}. Besides the existence of charge
pairing, the inhomogeneities of possible electronic nature can
exist in a form of spatially separated magnetic phases in cuprates
and manganites under doping
~\cite{
Kivelson_Lin}.
The magnetic inhomogeneities seen in other transition metal oxides
at the nanoscale level, widely discussed in the
literature~\cite{Dagotto,Nagaev,Nagaev1,Kiryukhin,Khomskii}, can
be crucial for the spin pairing instabilities, origin of
ferromagnetism and ferroelectricity in the spin and charge
subsystems~\cite{Nagaoka,Sokoloff}. A phase separation of the
ferromagnetic clusters embedded in an insulating matrix is
believed to be essential to the colossal
magnetoresistance (CMR)
in manganese oxides.
 At sufficiently low temperatures,
the spin redistribution in an ensemble of clusters can produce
inhomogeneities in the ground state and at finite
temperatures~\cite{cond-mat}.  The non-monotonous behavior of the
chemical potential versus electron concentration found in
generalized self consistent approximation~\cite{self} also
suggests possible electron instabilities and inhomogeneities near
half filling. From this perspective, exact studies at $T\geq 0$ of
electron charge and spin instabilities at various $U\geq 0$,
inter-site couplings
and various cluster topologies can give important clues for
understanding of charge/spin inhomogeneities and local
deformations for the mechanism of pairings and magnetism in
``large" concentrated systems whenever correlations are local.

It is a generally believed that a strong on-site Coulomb
interaction supports ferromagnetism and is detrimental for the
electron pairing and superconductivity in clusters and ``large"
concentrated systems~\cite{Kubo}. Our exact studies of gaps and
pseudogaps in finite-size systems have uncovered some important
answers related to spin-charge separation, pairing and thermal
condensation of the electron charge and spin. Despite this, there
is still a vast amount of uncertainties that need to be
unravelled: (i) What are the conditions for the electron phase
separation instabilities and spin/charge inhomogeneities? (ii)
What
is
the role of inhomogeneities and are these spatial spin/charge {\it
inhomogeneities} crucial for the pairing mechanisms in these
compounds? (iii) When treated exactly, what essential features can
the Hubbard clusters capture that share similar properties with
the ``large" concentrated transition metal oxides?

A redistribution of excess electron/hole inhomogeneities or spin
up/spin down domains in an ensemble of tetrahedrons for all $U>0$
depends on the sign of the hopping term~\cite{cond-mat}. Here we
show that in the distorted square pyramids 
in the perovskite structures, the inter-site coupling $c$ between
the apex site with the base can be beneficial or detrimental for
the electron pairing or ferromagnetism. An unstable ``saturated
ferromagnetism", existing in frustrated lattices at low
temperatures and large $U$ for a particular sign of hopping
($t>0$)~\cite{Nagaoka}, implies either antiferromagnetism,
unsaturated ferromagnetism, or electron coherent pairing for
charge and spin pairing (pseudo)gaps. Here it is argued that for
one hole off half filling electrons undergo separate thermal
condensation of the charge and spin degrees (independent of
cluster topology); the system may be divided into two coexisting
and dynamically bound bosonic subsystems, where two
types
of individual bosonic pairs, made up of double electron charges
and oppositely oriented (antiparallel) spins, can fluctuate. We
shall see that the phase diagram, under some circumstances, is
mostly controlled by the changes in the cluster geometry
(topology).

\section{Model and formalism}~\label{formalism}

It is possible to assume that the electron pairing and magnetic
instabilities of the purely electronic nature is described by a
local Coulomb interaction $U$ in a single band Hubbard model
\begin{eqnarray}
H=-t\!\sum\limits_{\left\langle i,\, j\right\rangle ,\, \sigma }
{c_{i\sigma }^+c_{j\sigma }}+ U\sum\limits_i {c_{i\uparrow }^+
c_{i\downarrow }^+c_{i\downarrow }c_{i\uparrow }}\label{2-site1}
\end{eqnarray}
The sign of the hopping amplitude $t$ between the nearest neighbor
sites in (\ref{2-site1})
leads
 to essential changes of electronic
structure. In nonbipartite clusters, such as tetrahedron, we
consider$t=\pm 1$. In addition, for the distorted pyramid we take
the coupling parameter between the apical site and the atoms in
the base equal to $ct$, with $c\leq 1$. Our studies of the quantum
and thermal fluctuations of electrons in finite clusters are based
on exact diagonalization, analytical and numerical calculations of
energy levels and expressions for the canonical and grand
canonical partition functions in various cluster geometries. The
exact grand canonical potential $\Omega_U$ for the interacting
electrons ($U$) in an external
 magnetic field ($h$) is
\begin{eqnarray}
\Omega_U=-{T}\ln \!\sum\limits_{n} e^{{-\frac {E_{n}-\mu N_{n} -
hs^{z}_n} T} }, \label{OmegaU}
\end{eqnarray}
where $N$ and $s^{z}$ are the number of particles and the
projection of the spin in the n-th quantum state. The first and
second order responses of the charge and spin degrees due to the
changes in the chemical potential $\mu$ (doping) or an applied
magnetic field are calculated without taking the thermodynamic
limit. The competing energy states, in conjunction with the
canonical and the grand canonical ensemble, yield valuable insight
into electron instabilities in the real nanoclusters and
nanomaterials with the correlated electrons. The introduced
formalism allows us to describe the smooth and sharp phase
transitions with competing stable and unstable inhomogeneities in
the
 canonical and grand canonical ensembles.

Below in Sec.~\ref{electron} we provide a detailed description of
the general methodology: we define the criteria for the charge and
the spin pairing instabilities in the canonical and grand
canonical ensembles; formulate the conditions for existence of
quantum critical points, coherent pairings and spontaneous
transitions in the ground state and corresponding critical
temperatures of crossovers for various phases and boundaries in
the phase diagrams discussed in Secs.~\ref{ground} and
\ref{temperature}.

\section{General methodology}~\label{electron}
\subsection{Canonical charge and spin gaps}~\label{gaps}
To facilitate the comparison with the frustrated clusters, we
summarize here the main results in the ground state and at the
finite temperatures for bipartite and nonbipartite clusters
obtained earlier in
Refs.~\cite{JMMM,PRB,PLA,PRL,computer,cond-mat,to be published}.
The degrees of freedom for charge and spin, electron and spin
pairings, temperature crossovers, quantum critical points, etc.
were extracted directly from the thermodynamics of these clusters.
One can classify the charge and spin order parameters as an energy
difference between the various competing phases by analogy with
phase transitions in the thermodynamic limit. In the ground state,
the calculated differences in the canonical energy levels between
configurations with various numbers of electron charge and spin
determine the energy gaps for electron charge and spin
excitations. Using the exact partition function in the canonical
ensemble, we also analyzed analytical expressions for the average
energies for various number of electrons $N$. For given
temperature $T$ and $U$, we calculated the energy differences
$\mu_+=E(N+1)-E(N)$ and $\mu_-=E(N)-E(N-1)$ for the average
canonical energies $E(N)$ by adding or subtracting one electron
(charge) in the cluster for a given spin $S$. The energy
difference between the two consecutive excitation energies by
adding or subtracting electron can serve as a natural order
parameter in a canonical approach. Then the charge gap at finite
temperature can be written as
${\Delta^{c}}(T)={\mu_{+}}-{\mu_{-}}=E(N+1)+E(N-1)-2E(N)$. The
opening of the gap is a local correlation effect, and clearly does
not follow from long range order, as exemplified here. The
difference $\mu_{+}-\mu_{-}$ is somewhat similar to the difference
$I-A$ for a cluster, where $I$ is the ionization potential and $A$
the electron affinity. For a single ``impurity" at
 half filling and $T=0$, $I-A$ is equal to $U$, which represents a
screened local parameter $U$ in the Hubbard model~\cite{herring}
(\ref{2-site1}). Thus the gap picture is analogous to an
inter-configuration energy gap for the crossover between different
many body ground state ionic configurations in solids. For
example, the charge gap is simply equivalent to the energy of the
``reaction" between different cluster configurations ($d$) at
fixed $N$
\begin{eqnarray}
d^N + d^N \to d^{N+1} + d^{N-1}, \label{reaction}
\end{eqnarray}
{\it i.e.}, the difference in the canonical energies of ionization
and affinity for many body cluster configurations in ensemble.
However, the configurational change in the ensemble of isolated
clusters is supposedly due to the possible spontaneous
fluctuations in the electron numbers and electron redistribution
via a charge reservoir. The negative spin gap in the canonical
ensemble can be treated correspondingly. We calculate a spin gap
as the difference in the average energies between the two cluster
configurations with
 various spin $S$ states,
$\Delta^s(T)=E(S+1)-E(S)$, for $E(S)$ being respectively the
average canonical energy in the spin sector at fixed
$N$~\cite{Yang}.

\subsection{Charge and spin instabilities}
Many phenomena and phase transitions invoked in the approximate
treatments of
``large" concentrated systems are seen also in
the
exact analysis of pairing instabilities in the canonical ensemble
of the small clusters in
 thermodynamic equilibrium~\cite{JMMM,PRB,PLA,PRL,computer,cond-mat,to be
published}. As we shall see, in some circumstances,
 small changes of the external parameters can lead to
level crossing instabilities in
various electron configurations with the formation of
negative charge and spin gaps. Physically, a positive gap
manifests the phase stability and smooth crossover, while a
negative gap describes spontaneous transitions from one stationary
state to another. Instead of a full phase separation at
$\Delta^{c,s}<0$, the local inhomogeneities in the clusters can
provoke electron redistribution and quantum mixing of the various
charge and spin configurations.
In the presence of a negative gap, the
many-body ground state has
an
appreciable probability of being found in either of these
competing configurations. The collective particle excitations are
also reflected in the fluctuations of the pair density in
Eq.~(\ref{reaction}). It is intriguing that these fluctuations
make the
pair
redistribution across the clusters possible even without direct
contact between the clusters. These fluctuations play a crucial
role of the pair transitions in the absence of electron hopping
between clusters in Eq.~(\ref{2-site1}). Near ground state
degeneracies, the lowest energy states control the low energy
dynamics of the electronic and magnetic transitions over a
significant portion of the phase diagram.

The possible quantum critical points, phase transitions and
nonzero temperature crossovers are described using a simple
cluster approach: we define critical parameters for the level
crossing degeneracies or quantum critical points from the
vanishing conditions for the canonical charge and spin gaps,
i.e., $\Delta^{c,s}(U,c)=0$. The sign of the gap is also important
in identifying the regions for the electron charge and spin
instabilities, such as the electron-electron $\Delta^c<0$,
electron-hole $\Delta^c>0$ pairings in the charge sector or the
parallel $\Delta^s<0$ and opposite $\Delta^s>0$ spin pairings in
the spin sector. The key question here is the exact relationship
between the canonical charge $\Delta^c$ gap and its corresponding
grand canonical spin $\Delta^s$ counterpart calculated for various
bipartite and frustrated cluster topologies. For charge degrees
the negative sign of gap implies phase (charge) separation ({\it
i.e., segregation}) of the clusters into hole-rich (charge
neutral) and hole-poor regions. The quantum mixing of the closely
degenerate, hole-poor $d^{N-1}$ and hole-rich $d^{N+1}$ clusters
for one hole off
half filling, instead of causing global phase separation, provides
a stable spatial inhomogeneous
medium that allows the pair charge to fluctuate. The
inhomogeneities favored by the negative gaps are essential for
providing the spontaneous redistribution of the electron charge or
spin.
The inhomogeneities in the charge redistribution for $\Delta^c<0$
and $\Delta^s=0$ imply static heterostructure for different
electron configurations, close in energy, in an unstable ensemble
of clusters. These inhomogeneities are consistent with nucleation
of the ``negative" charge gap in cuprates above
T$_c$~\cite{Yazdani}. At low temperatures, the dynamic picture for
pair fluctuations between different electron configurations
$\Delta^c<0$ and $\Delta^s>0$ is possible at relatively low
temperatures in spatially inhomogeneous coherent state
($\Delta^{s}\equiv-\Delta^c$, see also Sec.~\ref{squares}). This
result is consistent with the observation of nonlocal
superconductivity at low excitation energies and at higher
energies, holes localized in an inhomogeneous ``stripe"
pattern~\cite{Davis1}.
The negative spin gap describes the possible parallel spin pair
binding instability. This picture implies
spontaneous ferromagnetism and phase (spin) separation into
domains in accordance with the Nagaoka theorem. For the negative
gaps, one can introduce the critical temperatures $T^P_c(\mu)$ and
$T_{s}^F(\mu)$ versus chemical potential for boundaries between
various phases derived from the condition that
the
corresponding gaps disappear, i.e., $\Delta^{c,s}(T,\mu)=0$.

\subsection{Charge and spin susceptibility peaks}~\label{peaks}
Conventional phase transitions at finite temperature are driven by
thermal fluctuations. In
the
 grand canonical approach using exact analytical expressions
for the grand canonical potential and partition functions
 as expressed
 in Eq.~(\ref{OmegaU}), we
have
 analyzed (in
Refs.~\cite{JMMM,PRB,PLA,PRL,computer,cond-mat,to be published})
the variation of the charge, ${{\frac {\partial N} {\partial
\mu}}}$, and spin, ${{\frac {\partial s} {\partial h}}}$, density
of states
or corresponding charge $\chi_c(\mu)$ and spin $\chi_s(\mu)$
susceptibilities,
\begin{eqnarray}
{\chi_c}={{\frac {\partial {\left\langle {N}\right\rangle}}
{\partial \mu}}}, \ \ \ \ \ \ \ \ \chi_s={{\frac {\partial
\left\langle s^{z}\right\rangle} {\partial h}}}\label{spin}
\end{eqnarray}
as a function of the chemical potential $\mu$ and $h$ in
a
wide range of temperatures. In a grand canonical approach the
energy difference between the two consecutive susceptibility peaks
in terms of $\mu$ and $h$ at finite temperatures can serve as a
natural order parameters for charge and spin degrees respectively.
This energy difference for density of states in $\mu$ space
determines the charge gap in canonical approach. We find
(opposite) spin pairing gap by calculating the minimal magnetic
field necessary to overturn the spin. In the grand canonical
method we define the gap as a magnetic field at which the distance
between the subsequent spin susceptibility peaks in $\mu$ space
vanishes.
Using the maxima of zero magnetic field susceptibility, ${{\frac
{\partial s} {\partial h}}}|_{h\to 0}$, we also calculated the
boundary curve for the onset of the spin gap for various $\mu$ in
infinitesimal $h\to 0$ above $T_s^P$. To distinguish this from the
canonical and grand canonical gaps at finite temperatures we call
it {\sl pseudogap}.
The opening of such distinct and separated (pseudo)gap regions for
the spin and charge degrees at
various fillings in $\mu$ space is indicative
of the
 corresponding spin-charge separation. The crossover
temperatures and phase boundaries for various transitions can be
found by monitoring maxima and minima in charge and spin
susceptibilities. We define the critical temperatures $T_{c}$ and
$T^*$ in equilibrium as the temperature at which the distances
between the charge or spin susceptibility peaks vanish and
corresponding pseudogaps disappear (see Sec.~\ref{diagram}).
Notice that according to the given definition, the energy
pseudogaps obtained in the grand canonical method are positive
which is a key difference from the canonical gaps.

\begin{figure} 
\begin{center}
\includegraphics*[width=20pc]{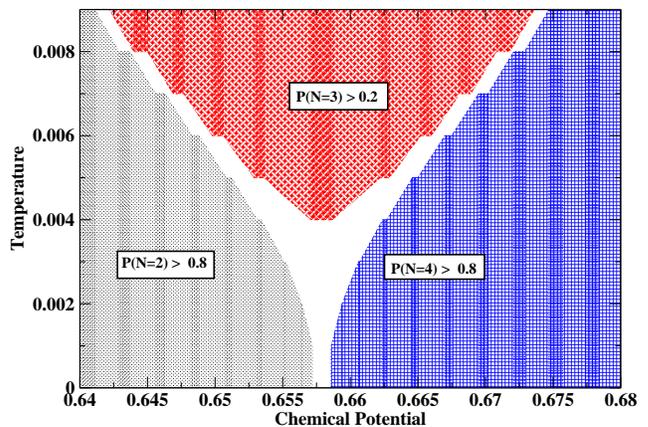}
\end{center}
\caption {Thermal occupation probabilities versus $\mu$ in the
grand canonical ensemble of 4-site clusters in the vicinity of
quantum critical point $\mu_P$ for $\left\langle
N\right\rangle\approx 3$ at $U=4.0$. The phase separation at
relatively low temperatures below $T\leq 0.0075$ manifests a
significant  suppression of $\left\langle N\right\rangle=3$
clusters close to optimal doping $\mu_P=6.557$. In this area
$\left\langle N\right\rangle=2$ and $\left\langle
N\right\rangle=4$ clusters share equal weight probabilities, while
at higher temperatures $\left\langle N\right\rangle\approx 3$ also
becomes thermodynamically stable. In equilibrium, the grand
canonical value $\mu_P$ at optimal doping at $T=0$ reproduces the
result $\mu_P=({\mu_++\mu_-)/2}$ for the canonical approach.}
\label{fig:num_mag}
\end{figure}

\subsection{Charge and spin inhomogeneities}~\label{inhomo}
The developed grand canonical approach can be applied to
understand of the electron fluctuations and the spatial
inhomogeneities to model the behavior of the concentrated systems
in bipartite and frustrated structures. An ensemble of bipartite
clusters at small and moderate $U$ exhibits typical inhomogeneous
behavior in its charge distribution. A normalized probability
$\omega_{N}$ for the electron distribution in grand canonical
ensemble as a function of temperature $T$ for various electron
numbers $N$ is the following
\begin{eqnarray}
\omega_{N}=\sum\limits_{n} e^{{-\frac {E_{nN}-\mu N } T}}/
\sum\limits_{n,N} e^{{-\frac {E_{nN}-\mu N} T}}.
 \label{OmegaU1}
\end{eqnarray}
The calculated probabilities of electrons in competing
configurations are shown in Fig.~\ref{fig:num_mag} for the 4-site
cluster at $U=4$. At low temperatures and electron concentration
close to $\mu_P$, the clusters with $\left\langle
N\right\rangle=2$ and $\left\langle N\right\rangle=4$ have equal
probabilities, $\omega_{2}=\omega_{4}\approx 0.5$. In some
circumstances electron configurations in equilibrium can have
close energies for the clusters in contact with a particle
reservoir. This picture shows a mixture of ungapped and partially
gapped states. As temperature increases, the probability
$\omega_{3}$ of $\left\langle N\right\rangle=3$ clusters with
unpaired spin gradually increases, while the probability of
finding spin paired, hole-rich and hole-poor clusters decreases.

Qualitatively, the formation of inhomogeneous electron
distribution or ``stripe" picture can be understood from simple
energy considerations~(see Sec.\ref{formalism}). For a fixed
average number of electrons, the charge and spin on each separate
cluster in the ensemble can fluctuate. The two configurations
close in energy are nearly degenerate, and, as temperature
increases, it is energetically favorable to have some clusters
with $d^{N-1}$ and another with $d^{N+1}$, instead of having
clusters with $d^{N}$ electrons. These results, that depend on the
cluster geometries, parameter $U$ as well as on the sign of $t$,
can be directly applied to nano and heterostructured materials,
which usually contain many independent clusters, weakly
interacting with one another with the possibility of having
inhomogeneities for a different number of electrons per cluster.
At half filling, the antiferromagnetic state has the lowest energy
per electron. Therefore, the energy can be minimized upon small
doping by
segregration
 of holes into charged clusters with different number
of electrons. The embedded antiferromagnetic background with
opposite spin pairing provides a spin rigidity (unperturbed) media
that allows inhomogeneities to optimize the coherent
pair fluctuations across the clusters~\cite{cond-mat}. The mixture
of the closely degenerate ferromagnetic domains can also lead to
the stable spatial magnetic inhomogeneities for spin fluctuations.
Interestingly, the quantum and thermal fluctuations in the
canonical and grand canonical ensembles display ``checkerboard"
patterns~\cite{Boyer}, nanophase inhomogeneities~\cite{Tranquada}
and nucleation of pseudogaps driven by temperature seen recently
in nanometer and atomic scale measurements~\cite{Hudson} in HTSCs
above $T_s^P$~\cite{Yazdani,Yazdani1}. Microscopic spatial
inhomogeneities and incoherent pairing pseudogaps in nanophases
measured by scanning tunnelling microscopy (STM) correlate
remarkably with our predictions using small 4-site and $2\times 4$
nanoclusters~\cite{PRL}.

\begin{figure} 
\begin{center}
\includegraphics*[width=20pc]{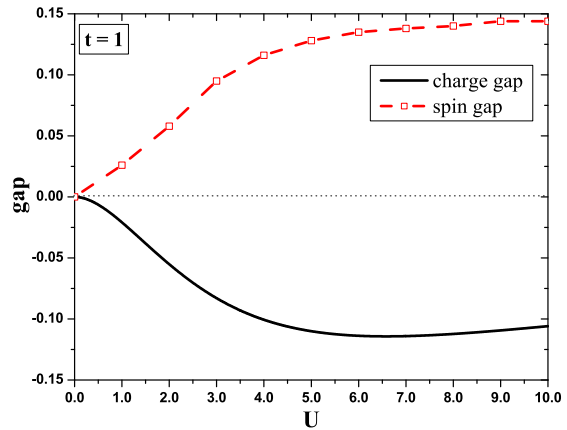}
\hfill
\end{center}
\caption {Charge $\Delta^{c}$ and spin $\Delta^{s}$ gaps versus
$U$ in an ensemble of tetrahedrons at $t=1$, $\left\langle
N\right\rangle\approx 3$ and $T=0.001$. Negative charge gap
$\Delta^{c}<0$ implies charge phase separation, while the
positive, opposite spin pairing gap of equal amplitude
$\Delta^s\equiv -\Delta^c$ describes Bose condensation of
electrons similar to BCS-like coherent pairing with a single
(unique) energy gap. This coherent state is analogous to Phase A
in 4-site clusters~\cite{cond-mat}. The spin gap has been
calculated using the grand canonical approach.}
\label{fig:gap-4-t=1 site}
\end{figure}
\begin{figure} 
\begin{center}
\includegraphics*[width=20pc]{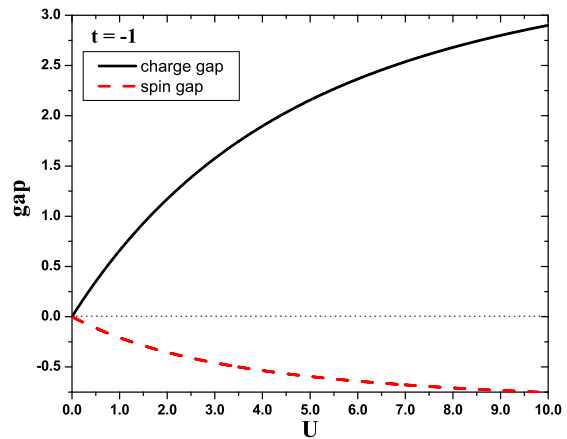}
\hfill
\end{center}
\caption {The charge $\Delta^{c}>0$ and parallel (triplet) spin
$\Delta^{s}<0$ gaps versus $U$ in an ensemble of tetrahedrons at
$t=-1$, $\left\langle N\right\rangle= 3$ and $T=0.001$. The
positive charge gap for all $U$ describes MH-like insulating
behavior analogous to Phase C in 4-site clusters~\cite{cond-mat}.
The negative spin gap ($\Delta^s<0$), coexisting with the charge
pairing gap displays $S={3\over 2}$ Nagaoka saturated
ferromagnetism at all $U>0$ (see Sec.~\ref{tetrahedron}). The
charge gap has been calculated using grand canonical approach.}
\label{fig:gap-4-t=-1 site}
\end{figure}

\subsection{Coherent charge and spin pairings}~\label{coherent}
The behavior of such clusters near crossing degeneracies in
a
quantum coherent phase with minimal spin at low temperatures is
somewhat similar to the conventional BCS superconductivity (see
Sec.~\ref{squares}). We found that at rather low temperatures the
calculated positive pseudospin gap $\Delta^s$ in the grand
canonical method can have equal amplitude with a negative charge
gap $\Delta^c$ derived in canonical method, $\Delta^s=|\Delta^c|$.
Such behavior is similar to the existence of a single gap in the
conventional BCS state. We call such an opposite spin (singlet)
coupling and electron charge pairing as a spin coherent electron
pairing in Ref.~\cite{cond-mat}. However, unlike to the BCS
theory, the charge gap differs significantly from the spin
pseudogap as temperature increases above $T_s^P$. For example,
the
vanishing of double peak structure in zero spin susceptibility
gives a critical temperature $T_s^P$, at which the spin pseudogap
disappear. The canonical charge gap disappears at higher
temperatures, i.e., $\Delta^c(T_c^P)=0$. The BCS-like coherent
behavior and possible superconductivity with condensation of
opposite spin pairs occur at rather low temperatures (see
Sec.~\ref{peaks}), while electron charge pairing can be
established at relatively
high
temperatures, $T_s^P<T_c^P$. The positive spin gap calculated in
grand canonical approach implies homogeneous electron (opposite)
spin spatial distribution below $T_s^P$. This picture is
consistent with the spatially homogeneous spin pseudogap that
opens below T$_c\equiv T_s^P$ in doping dependent STM measurements
of Bi$_2$Sr$_2$CuO$_{6+x}$~\cite{Boyer}. We
also find
a close analogy for the coherent electron pairing in clusters with
real space singlet pairs in resonance valence bond states or local
inter configuration fluctuations in mixed valence
states~\cite{Anderson,Anderson1,Anderson2,Hekking,Koch,AKOCHA}.

\section{Ground state properties}~\label{ground}
\subsection{Bipartite clusters}~\label{squares}
Exact calculations for charge and spin gaps in small clusters in
various geometries are important for understanding the electron
ground state behavior in bipartite and nonbipartite (frustrated)
systems. Below we summarize the results for electron instabilities
and phases obtained earlier (see Fig.~1 in Ref.~\cite{cond-mat})
for square and other bipartite clusters with one hole off half
filling at infinitesimal $T\to 0$.
The
vanishing of gaps at quantum critical points, $U_c=4.584$ and
$U_F=18.583$, indicates energy level crossings and electron
instabilities in 4-site clusters for charge and spin,
respectively. The charge $\Delta^{c}$ and spin $\Delta^{s}$ gaps
versus $U$ in an ensemble of square clusters at $\left\langle
N\right\rangle\approx 3$ exhibit  at infinitesimal $T\to 0$ the
following phases;  Phase A: Charge and spin pairing gaps of equal
amplitude $\Delta^{s}\equiv\Delta^{P}=-\Delta^c$ at $U\leq U_c$
describe Bose condensation of electrons similar to BCS-like
coherent pairing with a single energy gap; Phase B: Mott-Hubbard
like insulator with $\Delta^c>0$ and gapless $S={1\over 2}$
excitations at $U_c<U<U_F$ describes a spin liquid behavior; Phase
C: Parallel (triplet) spin pairing ($\Delta^s<0$) displays
$S={3\over 2}$ the saturated ferromagnetism at $U>U_F$ in
Mott-Hubbard insulator for a positive charge gap, $\Delta^c>0$.
Notice, that incoherent opposite spin pairing $|\Delta^s|\neq
\Delta^{c}$, different from the charge pairing at $U<U_c$,
suggests spin-charge separation for spin and charge degrees at
$U>U_F$.

Square clusters at weak and strong couplings share common
important features with 2${\times}$4 ladders and other bipartite
clusters~\cite{Nagaoka}.
Negative gaps describe possible hole binding or parallel spin
pairing instabilities. For charge degrees at weak coupling, this
gives an indication of phase separation ({\it i.e., segregation})
into hole-rich (charge neutral) and hole-poor clusters. In
contrast, at strong coupling the negative spin pairing gap for
parallel spins and positive charge gap reveal ferromagnetic
instability in accordance with the Nagaoka theorem. In large
bipartite clusters at intermediate $U$,
electrons behave differently from square clusters. For example, in
2${\times}$4 ladders we found an oscillatory behavior of charge
gap as a function of $U$~\cite{PRL}. The vanishing of the charge
gaps, manifesting the multiple level crossing degeneracies and
electronic instabilities in charge and spin sectors, is indicative
of possible electron instabilities in bipartite clusters at
moderate $U$.

\begin{figure} 
\begin{center}
\includegraphics*[width=15pc]{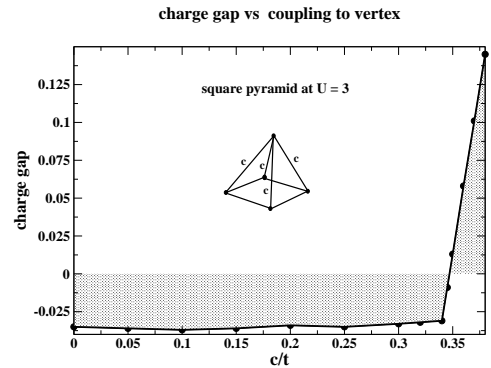}
\hfill
\end{center}
\caption {Charge gap $\Delta^c$ versus coupling $c$ between the
apex atom and the four base atoms in deformed square pyramid
($t=1$) for one hole of half filling, $\left\langle
N\right\rangle=4$, $U=3$ and $T=0.01$. Charge and spin pairing
gaps of equal amplitude $\Delta^{s}\equiv\Delta^{P}=-\Delta^c$ at
$c\leq 0.35$ imply coherent pairing, while $\Delta^c>0$ and
$\Delta^{s}<0$ at $c\geq 0.35$ correspond to a ferromagnetic
insulator for $S={1\over 2}$.}\label{fig:pyramid}
\end{figure}

\subsection{Tetrahedrons}~\label{tetrahedron}
For comparison with small bipartite clusters in
Sec.~\ref{squares}, we consider here a minimal four site
nonbipartite structure. A tetrahedron has a topology equivalent to
that of a square with the next nearest neighbor coupling
($t^\prime=t$) and may be regarded as a primitive unit of typical
frustrated system. Nonbipartite systems, without electron-hole
symmetry, exhibit
a pairing instability that depends on the sign of $t$. Notice that
sign of $t$ also
leads
to essential changes in the electronic structure. The tetrahedral
clusters show pairing instabilities for charge degrees at $t=1$
and spin degrees at $t=-1$ that maximizes the amplitudes of
negative charge $\Delta^c<0$ and spin $\Delta^s<0$ gaps and
corresponding condensation temperatures, $T^P_c(\mu)$ and
$T_{s}^F(\mu)$~\cite{Serg}. The negative gap in the canonical
approach displays electron pairing $\Delta^P=|\Delta^c|$
instability for all $U$. Fig.~\ref{fig:gap-4-t=-1 site}
illustrates the charge and spin gaps at small and moderate $U$.
The negative charge gap in Fig.~\ref{fig:gap-4-t=1 site} is
indicative of the inhomogeneous charge redistribution and phase
separation of electron charge into hole-rich (charged) and
hole-poor (neutral) cluster configurations~\cite{PRB}. The phase
diagram for $t=1$ is similar to the Phase A in Sec.~\ref{squares},
but applied for all $U$ values. In contrast, the positive spin gap
in the grand canonical approach $\Delta^s>0$ corresponds to
uniform opposite spin distribution in Fig.~\ref{fig:gap-4-t=1
site}. This BCS-like picture for charge and spin gaps of equal
amplitude $\Delta^{s}\equiv\Delta^{P}=-\Delta^c$ at $\left\langle
N\right\rangle \approx 3$, in analogy with the square clusters,
will be called coherent pairing (CP)~\cite{cond-mat}. In
equilibrium, the spin singlet background ($\chi_s>0$) stabilizes
phase separation of paired electron charge in a quantum CP phase.
Fig.~\ref{fig:gap-4-t=1 site} illustrates the charge $\Delta^c$
and spin $\Delta^s$ gaps in tetrahedral clusters at $\left\langle
N\right\rangle \approx 3$, $T\to 0$. The unique gap, $\Delta^s
\equiv\Delta^P$ at $T=0$, in Fig.~\ref{fig:gap-4-t=1 site} is
consistent with the existence of a single quasiparticle energy gap
in the BCS theory for $U<0$~\cite{to be published}. Positive spin
gap for all $U$ provides pair rigidity in response to a magnetic
field and temperature (see Sec.~\ref{diagram}). Notice that the
coherent pairing exists also at large $U$ where Nagaoka theorem
for nonbipartite clusters with specific sign of $t$ can be
applied. The stability of minimal spin $S=0$ (singlet) state in
tetrahedron at $t=1$ is consistent with
that of non maximum (unsaturated) spin in Nagaoka problem.
Thus our result shows that Nagaoka instability toward spin flip at
large $U$ in frustrated lattices with $t=1$ can be associated with
the BCS-like coherent
pairing applied for general $U$.

The negative spin gap $\Delta^s$ in Fig.~\ref{fig:gap-4-t=-1 site}
is shown for canonical energy differences between $S={3\over 2}$
and $S={1\over 2}$ configurations. Correspondingly, the positive
charge gap $\Delta^c>0$ in a stable MH-like state is derived using
grand canonical energies~\cite{PRB}. As in bipartite square
clusters, the grand canonical positive charge gap $\Delta^c>0$ is
different (incoherent) from the parallel spin pairing gap,
$\Delta^s<0$. Thus the phase diagram for $t=-1$ with
$\Delta^{c}\neq -\Delta^s$ is similar to the Phase C in
Sec.~\ref{squares}, but applied for all $U$ values in the phase
diagram. The negative spin gap for all couplings implies parallel
spin pairing and
Nagaoka-like saturated ferromagnetism with maximum spin in the
entire range of $U$. Spin-charge separation is considered to be
one of the key properties of the correlated electrons that
distinguishes $t=-1$ from $t=1$. Such behavior at $t=-1$ is
accompanied by spin-charge separation and formation of the
magnetic (spatial) inhomogeneities or domain structures~\cite{PRB}
in a wide range of parameters.

\subsection{Square pyramids}
From the early days of high-temperature superconductivity, the
idea of a possible role of apical sites in p-type superconductors
has been controversial. The oxygen atom position at the apex of
pyramidal crystalline structure can be altered through the
addition of impurities and can be relocated to a lower or sideways
position, thus changing the electron interactions or coupling
strength $c$ between apex and the planar atoms.  There is no
significant influence of localized electron charge of apical site
on electron pairing and possible superconductivity in CuO$_2$
planes in Bi$_2$Sr$_2$CaCu$_2$O$_{8+\delta}$. When excess apex
does not exist, {\it i.e.}, $\delta=0$, this system is
an
 insulator.
However, when excess apex oxygen is introduced, hole carriers are
supplied into CuO$_2$ planes and
the
 material shows
superconductivity~\cite{Book}. Here we try to draw a closer
connection to HTSCs perovskites and consider an ensemble of square
pyramids of octahedral structure.

Fig.~\ref{fig:pyramid} shows the charge gap at fixed $U=3$ and
$\left\langle N\right\rangle \approx 4$ under the variation of the
coupling term $c$ between the plane and the apex atoms. This
picture gives surprisingly plausible evidence for understanding
the detrimental role of excess electron on charge pairing for
possible distortions of pyramidal crystalline structure in
perovskites.  In Fig.~\ref{fig:pyramid}, the strong distortion of
the pyramid structure for $c=0$ (with reduced coordination number)
reproduces a charge pairing gap in planar square geometries. At
$\left\langle N\right\rangle \approx 4$, the electron is localized
and there is no charge transfer from apex atom in an ensemble of
pyramid clusters at $c=0$. The negative charge gap, identical to
the spin gap, exists only for $c\leq c_0$, where $c_0= 0.35$ is a
quantum critical point for level crossing degeneracy. Calculated
electron distribution, as a function of $c$, shows that electron
charge residing on the apical site does not contribute
 to the
pairing whenever $c$ is less than $c_0$. The coupling in the
pyramid structure at $c<c_0$ for $\left\langle N\right\rangle
\approx 4$ leads to charge pairing instability with negative
charge and positive spin gaps of equal amplitude as seen in square
clusters at $\left\langle N\right\rangle \approx 3$ in
Sec.~\ref{squares}. In contrast, at $c>c_0$, the induced charge
gap driven by $c$ change leads to
 electron hole pairing and a
transition into insulating Mott-Hubbard (MH) behavior with
$\Delta^c>0$.
The apex atom, coupled to square-planar geometry, have shown to
have a detrimental affect on the negative charge and positive spin
gaps, which are favorable to forming a Bose condensate in the
region of instability. We found a coherent pairing in the phase
diagram with one hole off half filling also in the ensemble of
octahedron clusters (perovskite systems) in Ref.~\cite{Fernando1}.
There is also shown that octahedron threaded by magnetic flux in
hole-rich regions can get trapped in stable minima at half
integral units of the magnetic quantum flux.
Such approach can be applied to understand the detrimental effect
of the transverse magnetic field on electron charge and opposite
spin pairings for possible superconductivity in HTSCs in planar
face centered square (fcs) geometry~\cite{cond-mat}.

\begin{figure} 
\begin{center}
\includegraphics*[width=20pc]{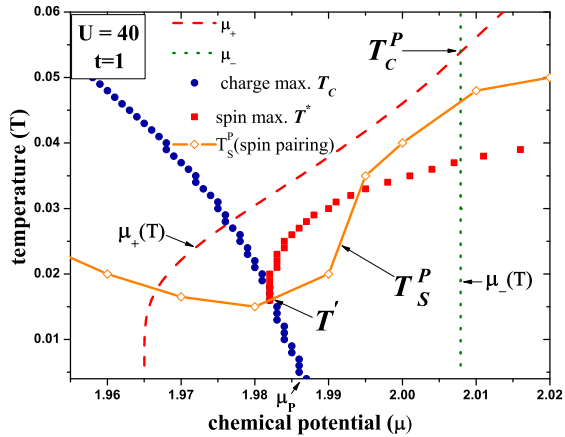}
\hfill
\end{center}
\caption {The $T$-$\mu$ phase diagram of tetrahedrons without
electron-hole symmetry at optimally doped $\left\langle
N\right\rangle\approx 3$ regime near $\mu_P=1.998$ at $U=40$ and
$t=1$ illustrates the condensation of electron charge and onset of
phase separation for charge degrees below $T_c^P$. The incoherent
phase of preformed pairs with unpaired opposite spins exists above
${T_s}^P$. Below $T_s^P$, the paired spin and charge coexist in a
coherent pairing phase. The charge and spin susceptibility peaks,
denoted by $T^*$ and $T_c$, define pseudogap regions calculated in
the grand canonical ensemble, while phase boundaries $\mu_+(T)$
and $\mu_-(T)$ are evaluated in the canonical ensemble. The spin
pseudogap region exists for $T_s^P<T<T^{\prime}$. Charge and spin
peaks reconcile at $T\sim T^\prime$, while $\chi^c$ peak below
$T_s^P$ signifies metallic (charge) liquid (see inset for square
cluster in Ref.~\cite{PRL}).}\label{fig:ph4_6}
\end{figure}

\section{Phase T-$\mu$ diagram}~\label{temperature}
\subsection{Tetrahedrons at large $U$ and $t=1$}~\label{diagram}
The charge and spin susceptibility peaks in clusters, reminiscent
of the singularities in infinite systems, display an extremely
rich phase diagram at finite temperatures. The realization of a
high transition temperature, T$_c$, in clusters and bulk systems
depends on the interaction strength $U$, doping,
 and
 the detailed
nature of the crystal structure (sign and amplitude of $t$). As
exemplified here, the critical temperatures for various pairing
instabilities in frustrated clusters also strongly depend on the
sign of the hopping ($t$) term. Fig.~\ref{fig:ph4_6} for $t=1$
illustrates a number of nanophases, defined in
Refs.~\cite{PRB,PRL}, for the tetrahedron at large $U=40$, found
earlier in tetrahedron and bipartite 2${\times}$2 and 2${\times}$4
clusters at moderate $U=4$ values~\cite{cond-mat}. This diagram
 captures the essential electron charge and spin pairing
instabilities at finite temperatures. The curve $\mu_+(T)$ below
$T_c^P$ signifies the onset of charge pair condensation. The
calculated susceptibility peaks in Fig.~\ref{fig:ph4_6} correspond
to the pseudogap crossover temperature $T^*$. As temperature is
lowered below $T^*$, a spin pseudogap is opened up first, as seen
in NMR experiments~\cite{PRL}, followed by the gradual
disappearance of the spin excitations, consistent with the
suppression of low-energy excitations in the HTSCs probed by STM
and ARPES~\cite{Bodi,hashini,Boyer,Yazdani,Yazdani1}.
In contrast, the local charge gap, $\Delta^c$, evolves smoothly as
temperature decreases below $T_c^P$.
The opposite spin CP phase, with fully gapped collective
excitations, begins to form at $T\leq {T_s}^P$ and spin pairing
rigidity gradually grows upon lowering of the temperature. As
temperature decreases both charge and spin pseudogaps emerge
into
 one gap at zero temperature.
Therefore, at sufficiently low temperatures, this leads to the
BCS-like coherent coupling of electron charge to bosonic
excitations~(see Sec.~\ref{tetrahedron}). However, the spin gap is
more fragile and as temperature increases it vanishes at
${T_s}^P$, while charge pseudogap survives until $T_c^P$.

The charge inhomogeneities~\cite{Davis1,Valla} in hole-rich and
charge neutral {\it spinodal} regions between $\mu_+$ and $\mu_-$
are similar to those found in the ensemble of squares and resemble
important features seen in the HTSCs. Pairing and transfer of
holes is a consequence of the existence of an inhomogeneous
background. In the absence of direct contact between clusters, the
inhomogeneities in the grand canonical approach are establishing a
transfer of paired electrons via this (thermal) bath media.
Fig.~\ref{fig:ph4_6} shows the presence of bosonic modes below
$\mu_+(T)$ and $T_s^P$ for paired electron charge and opposite
spin respectively. This picture suggests condensation of electron
charge and spin at various crossover temperatures while
condensation in the BCS theory occurs at a unique T$_c$ value.
This result suggests that thermal excitations in the exact
solution are not quasiparticle-like renormalized electrons, as in
the BCS theory, but collective paired charge and coupled opposite
spins~\cite{cond-mat}.

The coherent pairing of holes here is a consequence of the
existence of homogeneous opposite spin pairing background,
consistent with the STM measurements~\cite{Hudson}. This led us to
conclude that $T_s^P$ can be relevant to the superconducting
condensation temperature T$_c$ in the HTSCs. In the absence of
spin pairing above $T_s^P$, the pair fluctuations between the two
lowest energy states becomes incoherent. The temperature driven
spin-charge separation above $T_{s}^P$ resembles an incoherent
pairing (IP) phase seen in the
HTSCs~\cite{Valla,Bodi,hashini,Boyer,Yazdani}. The charged pairs
without spin rigidity above $T_{s}^P$, instead of becoming
superconducting, coexist in a nonuniform, charge degenerate IP
state similar to a ferroelectric phase~\cite{to be published}. The
unpaired weak moment, induced by a field above $T_s^P$, agrees
with the observation of competing dormant magnetic states in the
HTSCs~\cite{hashini}. The coinciding $\chi^s$ and $\chi^c$ peaks
in the vicinity of critical temperature $T^\prime$ show full
reconciliation of charge and spin degrees seen in the HTSCs above
T$_c$. However, in both channels the charge and spin pseudogaps
behave differently or independently. Indeed, we find that the
variation of the spin pairing gap with temperature does not cause
a change in the charge pairing gap. In the absence of
electron-hole symmetry in the tetrahedrons, the reentrant
phenomenon can be observed at low temperatures~\cite{cond-mat}. In
Fig.~\ref{fig:ph4_6}, as temperature increases near optimal doping
$\mu\leq\mu_P$, clusters undergo a transition from a CP phase to a
MH-like behavior.
Notice that the charge and spin pairing do not disappear in the
underdoped regime for $\mu\geq \mu_P$ but are governed
predominantly by the physics of antiferromagnets at half filling.
In contrast, in the overdoped regime at low temperatures, the
charge pairing pseudogap gradually approaches
 the spin pseudogap as in the conventional BCS theory.

Our exact calculations of phase diagrams in various bipartite and
nonbipartite clusters provide strong evidence for the existence of
a narrow, homogeneous (pseudo)gap $\Delta^s$ that vanishes near
$T_s^P$, coexisting with inhomogeneous, weakly temperature
dependent broad gap $\Delta^c$, which disappears at higher
temperatures $T_c^P>T_s^P$. These phase diagrams display coherent
and incoherent pairing (pseudo)gaps and possible superconductivity
in agreement with the recent STM measurements in
HTSCs~\cite{Davis1,Valla,Bodi,hashini,Boyer,Yazdani,Yazdani1}.
\bigskip
\begin{figure} 
\begin{center}
\includegraphics*[width=20pc]{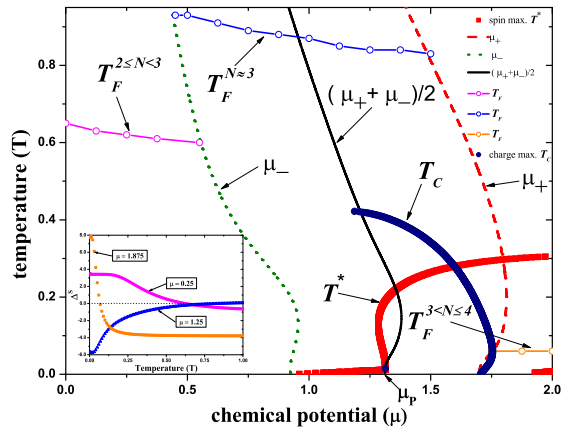}
\hfill
\end{center}
\caption { The $T$-$\mu$ phase diagram of square clusters near
optimally doped $\left\langle N\right\rangle\approx 3$ regime at
$U=40$ illustrates the condensation of electron spin and onset of
phase separation for spin degrees below ${T}^{N}_F$ for various
$N$ regions in $\mu$ space. The charge and spin susceptibility
peaks, denoted by $T_c$ and $T^*$, define corresponding pseudogap
regions calculated in the grand canonical ensemble, while
boundaries for stable ferromagnetic transitions $\mu_+(T)$ and
$\mu_-(T)$ are evaluated in the canonical ensemble. In
equilibrium, the grand canonical value $\mu_P$ at optimal doping
$\left\langle N\right\rangle=3$ at $T=0$ reproduces the result
$\mu_P=({\mu_++\mu_-)/2}$ for canonical approach. The inset shows
variation of canonical spin gap $\Delta^s$ versus temperature for
various values of $\mu$.}\label{fig:ph4_7}
\end{figure}

\subsection{Bipartite clusters at large $U$}~\label{diagram}
As for the cuprates, the bipartite clusters are useful for
understanding the magnetic behavior and instabilities in
manganites. The phase diagram in Fig.~\ref{fig:ph4_7} for square
clusters at $U=40$, quite similar to other bipartite clusters at
large $U$ limit~\cite{PLA}, displays characteristic features of
managanites with strong electron correlations. In the ground
state,
the
cluster at $\left\langle N\right\rangle= 3$ exhibits
ferromagnetism in agreement with the Nagaoka
theorem~\cite{Nagaoka}. However, we observe saturated
ferromagnetism, $S={3\over 2}$ spin state for $\left\langle
N\right\rangle\approx 3$ clusters with one hole off half filling
also at finite temperatures. The curve below ${T}^{N\approx3}_F$
signifies the onset of spontaneous magnetization with $S={3\over
2}$ for parallel spin condensation. The positive charge gap
($\Delta^c=0.7787$) for electron-hole (exciton) pairing manifests
MH-like insulating behavior. In contrast, clusters show the
minimum spin $S=0$ antiferromagnetism at $\left\langle
N\right\rangle\approx 2$ and $\left\langle N\right\rangle\approx
4$ in the ground state and at finite temperatures. The inset in
Fig.~\ref{fig:ph4_7} displays the variation of spin gap for
various regions. At $\mu=1.35$ the negative spin gap for
$\left\langle N\right\rangle\approx 3$ approaches zero as $T\to
{T}^{N\approx3}_F$. Thus the region above ${T}^{N\approx 3}_F$
describes
a
 paramagnetic phase with zero spin gap for unpaired
spins. In contrast, the positive spin gap in high doped regime at
$\mu=0.25$ changes its sign at temperatures above ${T}^{2\leq
N<3}_F$. This picture describes a transition driven by temperature
from antiferromagnetism into ferromagnetism with $S=1$. At half
filling, a MH-like antiferromagnetism is stable at very low
temperatures and unsaturated ferromagnetic state with $S=1$
becomes more stable at higher temperatures, $T\geq {T}^{3<
N\leq4}_F$. However, ${T}^{3< N\leq4}_F\to 0$ as $U\to \infty$ and
unsaturated ferromagnetism with $S=1$ at half filling can be
stabilized at infinitesimal temperatures. The well separated
charge and spin susceptibility curves in
the 
 entire parameter range
near $\left\langle N\right\rangle\approx 3$ show spin-charge
separation and decoupling of charge and spin degrees. The
susceptibility peak at $T^*$ for $\left\langle N\right\rangle
\approx 3$ in Fig.~\ref{fig:ph4_7} displays a spin liquid behavior
in the overdoped region for $\mu\leq \mu_P$, while well developed
negative spin gap in underdoped for $\mu>\mu_ P$ regions at low
temperatures describes a ferromagnetic insulator.
In Fig.~\ref{fig:ph4_7}, the region of metallic-like behavior is
manifested by the charge susceptibility peaks along the $T_c$
curve.

Phase diagram with $\mu$ dependent locally inhomogeneous,
$\Delta^s<0$, and homogeneous, $\Delta^s>0$, spin structures at
low temperatures, coexisting with charge ordered homogeneous
Mott-Hubbard like gap $\Delta^c>0$ displays spin-charge separation
and characteristic features of the CMR-manganite
La$_{1–x}$Ca$_x$MnO$_3$ and related materials with alternating
insulating ferromagnetic and charge ordered antiferromagnetic
regions~\cite{CMR}.

\section{Conclusion}
We
have
 studied the dependence of the ground state and thermal
properties in
the
repulsive Hubbard model on cluster size, geometry, electron number
and interaction strength to understand the inhomogeneous
superconducting elements and stripes. The inhomogeneities found in
exact calculations of clusters are promising for
the 
 description of geometric stripes with alternating
superconducting and antiferromagnetic regions in high-Tc cuprates
and magnetic domain structures in manganites. Spatial electron
inhomogeneities capture the magnetic and pairing instabilities in
clusters and respective bulk materials. The principal conclusion
is that the exact solution for optimal inhomogeneities mimicked in
small clusters can target essential features relevant to existing
inhomogeneities in nanostructured materials on a nanoscale level.
We found charge and spin gaps of equal amplitude in the ground
state similar to the coherent pairing in conventional BCS theory.
However, separate Bose condensation of electron charge and spin
degrees with two consecutive transition temperatures into coherent
pairing
 suggests a mechanism different from the
 prediction of the BCS coherent behavior with a
unique critical temperature. This picture is also consistent with
the existence of two different energy scales for electron charge
and spin pairing condensation temperatures in the
HTSCs~\cite{Davis1,Valla,Bodi,hashini,Boyer,Yazdani}.

The electronic instabilities in various geometries
and in a
 wide range of
$U$ and temperatures will be useful for the prediction of
coherent and incoherent
electron
 pairings, ferroelectricity~\cite{deHeer0,to be published} and possible
superconductivity in nanoparticles, doped cuprates, etc. In
contrast to bipartite clusters, the exact solution for the
tetrahedron depends on the sign of $t$ and shows relatively weak
dependence on $U$.
 For example, the tetrahedron exhibits similar
features at $U=4$ and $U=40$ whenever $t=1$. On other hand, the
behavior of the tetrahedron for $t=-1$ strongly differs from that
of $t=1$. These results for frustrated clusters show that the
properties are more sensitive to the change of the sign of the
hopping term rather than the $U$ parameter. This fact can explain
why itinerant ferromagnetism can occur even at relatively weak
interactions in frustrated systems. Our findings at small,
moderate and large $U$ carry a wealth of information regarding
phase separation, ferromagnetism and Nagaoka instabilities in
bipartite and frustrated nanostructures in manganites/CMR
materials at finite temperatures. These exact results allow us to
understand the origin of level crossings, spin-charge separation,
reconciliation and full Bose condensation~\cite{Lee}. The obtained
phase diagrams provide novel insight into electron condensation,
magnetism, ferroelectricity at finite temperatures and display a
number of inhomogeneous, coherent and incoherent nanophases seen
recently by STM and ARPES in numerous nanomaterials, assembled
nanoclusters and ultra-cold fermionic atoms~\cite{TM,Fermi}.

Finally, we
conclude
 that the use of the chemical potential and
the departure from zero degree singularities in the canonical and
grand canonical ensembles are essential for understanding
 the important thermal properties and physics of phase separation instabilities and
inhomogeneities at nanoscale level. The article currently in
progress is aimed to study the stability of pairing correlations
and magnetism in the presence of transverse magnetic
field~\cite{Fernando1}. It will be shown that magnetic flux tube
inside the octahedral cluster can get trapped in stable minima at
half integral units of the flux quantum in hole-rich regions.

We thank Daniil Khomskii and Valery Pokrovsky for helpful
discussions. This research was supported in part by U.S.
Department of Energy under Contract No. DE-AC02-98CH10886.


\begin{thebibliography}{0}
\bibitem{Davis1} Y. Kohsaka {\it et al}., Science {\bf 315} (2007) 1380.
\bibitem{Valla} T.~Valla {\it et al}., Science {\bf 314}
(2006) 1914.
\bibitem{Bodi} A.~C. B$\acute{o}$di, R.~Laiho, and E.~L$\ddot{a}$hderanta,
Physica C{\bf 411} (2004) 107.
\bibitem{hashini} H.~E.~Mohottala {\it et al}.,
Nature Materials~{\bf 5} (2006) 377.
\bibitem{Boyer} M. C. Boyer,
Nat. Phys. {\bf 3} (2007) 802.
\bibitem{Yazdani} K.~K.~Gomes {\it
et al}.,
Nature {\bf 447} (2007) 569.

\bibitem{Yazdani1} A.~N.~Pasupathy {\it et al}., Science {\bf 320} (2008) 196.
\bibitem{Cohen} R.~E. Cohen, Nature {\bf 358} (2005) 136.
\bibitem{deHeer0}
 R.~Moro, S.~Yin, X.~Xu, and W.~A. de~Heer, Phys.~Rev.~Lett. {\bf 93} (2004)
086803;
X.~Xu, S.~Yin, R.~Moro, and W.~A. de~Heer, {\bf 95}, (2005)
237209.
\bibitem{TM} S.~Y. Wang, J.~Z. Yu, H. Mizuseki, Q. Sun, C.~Y. Wang,
and Y. Kawazoe, Phys. Rev. B{\bf 70} (2004) 165413.
\bibitem{Yang} C.~Yang, A.~N.~Kocharian, and Y.~L.~Chiang,
J. Phys.: Condens. Matter B{\bf 12} (2000) 7433.
\bibitem{Shiba}
H.~Shiba and P.~A.~Pincus, Phys.~Rev. B{\bf 5} (1972) 1966.
\bibitem{Falicov} L.~M.~Falicov
and R.~H.~Victora , Phys.~Rev. B{\bf 30} (1984) 1695.
\bibitem{Calaway} L.~Tan and J.~Calaway, Phys.~Rev. B{\bf 46},
5499 (1992); ibid {\bf 35} (1987) 8723.
\bibitem{Schumann1} R.~Schumann, Ann.~Phys. {\bf 11} (2002) 49; 
%
{\it ibid}
{\bf 17} (2008) 221.
\bibitem{Hirsch} J.~H.~Hirsch, Phys.~Rev.~B{\bf 67} (2003) 035103.
\bibitem{pairing}
S. Belluci, M.Cini, P. Onorato, and E. Perfetto, J. Phys.:
Condens. Matter {\bf 18}, S2115 (2006); W.-F. Tsai and S.A.
Kivelson, Phys. Rev. B{\bf 73}, 214510 (2006); S.R. White, S.
Chakravarty, M.P. Gelfand, and S.A. Kivelson, {\it ibid} B{\bf
45}, 5062 (1992); R.M. Fye, M.J. Martins, and R.T. Scalettar, {\it
ibid} {\bf 42}, R6809 (1990); N.E. Bickers, D.J. Scalapino, and
R.T. Scalettar, Int. J. Mod. Phys. B{\bf 1}, 687 (1987).
\bibitem{arXiv} A.~N.~Kocharian,
G.~W.~Fernando, K.~Palandage, and J.~W.~Davenport,
arXiv:cond-mat.str-el/0510609v1 (2005) (unpublished).
\bibitem{JMMM} A.~N.~Kocharian, G.~W.~Fernando, K. Palandage and
J.~W.~Davenport, J.~Magn.~Magn.~Mater. {\bf 300} (2006) e585.
\bibitem{PRB} A.~N.~Kocharian,
G.~W.~Fernando, K.~Palandage, and J.~W.~Davenport, Phys.~Rev.
B{\bf 74} (2006) 024511.
\bibitem{PLA} A.~N.~Kocharian,
G.~W.~Fernando, K.~Palandage, and J.~W.~Davenport, Phys.~Lett.
A{\bf 364} (2007) 57.
\bibitem{PRL} G.~W.~Fernando,
A.~N.~Kocharian, K. Palandage, Tun Wang, and J.~W.~Davenport,
Phys.~Rev. B{\bf 75} (2007) 085109.
\bibitem{computer} K. Palandage, G.~W.~Fernando,
A.~N.~Kocharian and J.~W.~Davenport, J. Comput.-Aided. Mater. Des.
{\bf 14} (2007) 103.
\bibitem{cond-mat}
A.~N.~Kocharian, G.~W.~Fernando, K.~Palandage, and
J.~W.~Davenport,
Phys.~Rev. B{\bf 78} (2008) 075431.

\bibitem{to be published}
A.~N.~Kocharian, G.~W.~Fernando, K. Palandage, and
J.~W.~Davenport, ISPM Seattle'08 conference proceedings,
Ultramicroscopy, to be published (2009).
\bibitem{Tranquada} J.~M.~Tranquada {\it et al}.,
Nature (London) {\bf 375} (1995) 561.
\bibitem{Arigoni} E.~Arrigoni, and
S.~A.~Kivelson, Phys.~Rev. B{\bf 68} (2003) 180503.

\bibitem{Balatsky} J.~Eroles, G.~Ortiz, .A.~V.~Balatsky,
A.~R.~Bishop, Inter.~J.~Mod.~Phys. {\bf 15} (2001) 2833.
\bibitem{Hudson}
W. D. Wise {\it et al}.,
Nature Physics {\bf 4}, 696 (2008).
\bibitem{Kivelson_Lin} V. J. Emery, S. A. Kivelson, and H. Q. Lin,
Phys.~Rev.~Lett. {\bf 64} (1990) 475.
\bibitem{Dagotto} E. Dagotto, Science {\bf 309} (2005)
257.Phys.~Rev.~Lett. {\bf 64} (1990) 475.
\bibitem{Nagaev} E.~L.~Nagaev, Physics - Uspechi {\bf 39}
(1996) 781.
\bibitem{Nagaev1} E.~L.~Nagaev, Physica B{\bf 230-232} (1997) 816.
\bibitem{Kiryukhin} V. Kiryukhin, T.~Y. Koo, H.~Ishibashi, J.~P.~Hill, and S-W.~Cheong,
Phys.~Rev. B{\bf 67} (2003) 064421.
\bibitem{Khomskii} L.~N. Bulaevskii, C.~D. Batista,
M.~V. Mostovoy, D.~I. Khomskii, Phys.~Rev. B{\bf 78} (2008)
024402.
\bibitem{Nagaoka} Y.~Nagaoka, Phys.~Rev. {\bf 147} (1966) 392.
\bibitem{Sokoloff} J.~B.~Sokoloff, Phys.~Rev. B{\bf 3} (1971) 3826.
\bibitem{self} A.~N.~Kocharian,
C.~Yang, Y.~L.~Chiang, and T.~Y.~Chou, Inter. J. Mod.~Phys. B{\bf
17} (2003) 5749.
\bibitem{Kubo} W.~P.~Halperin, Rev.~Mod.~Phys. {\bf 58} (1986) 533.
\bibitem{herring} C. Herring, "Exchange Interactions Among Itinerant Electrons", in
Magnetism Vol IV, G.T. Rado and H. Suhl eds., Academic Press, New
York, 1966.

\bibitem{Anderson} P~W.~Anderson, Mater.~Res.~Bul. {\bf 8} (1973) 153.
\bibitem{Anderson1} P.~Fazekas and P~W.~Anderson,
Philos.~Mag. {\bf 30} (1974) 432.
\bibitem{Anderson2} P~W.~Anderson, Science {\bf 235} (1967) 1196.
\bibitem{Hekking} F.~W.~J.~Hekking et al., Phys. Rev. Lett. 70 (1993) 4138.
\bibitem{Koch} J.~Koch, M.~E.~Raikh, and F.~von~Oppen, Phys.~Rev.~Lett. {\bf 90} (2006) 056803.
\bibitem{AKOCHA} A.~N.~Kocharian and Khomskii,
[Zh. Eksp. Teor. Fiz., {\bf 71} (1976) 767] Sov. Phys. JETP {\bf
44} (1976) 404.
\bibitem{Serg} I.~A.~Sergienko and S.~H.~Curnoe,
Phys.~Rev. B{\bf 70} (2004) 144522.
\bibitem{Book}H. Kamimura, H. Ushio, S. Matsuno, T. Hamada, Theory of Copper
Oxide Superconductors, Chapter VII: Electronic Structure of a
CuO$_5$ Pyramid in Bi$_2$Sr$_2$CaCu$_2$O$_{8+d}$, Springer Berlin
Heidelberg, pp 51-53 (2005).


\bibitem{Fernando1} G.~W.~Fernando, K. Palandage,
A.~N.~Kocharian,and J.~W.~Davenport, submitted to Phys.~Rev. B
(2009).
\bibitem{Murakami} Y. Murakami {\it et al}.,
Nature {\bf 423} (2003) 965.
\bibitem{CMR} New Trends in the Characterization of CMR-Manganites and Related
Materials, edited by K. Baerner Research Signpost, Trivandrum
(2005).
\bibitem{Lee} R.~Friedberg, T.~D.~Lee, and H.~C.~Ren,
Phys.~Rev. B{\bf 50} (1994) 10190.
\bibitem{Fermi} J. K. Chin {\it et al}., Nature {\bf 443} (2006) 961.

\end{thebibliography}
\end{document}